\documentstyle[12pt,aaspp4]{article}
\def\h{Hipparcos }
\def\plei{Pleiades }

\def\diff{different }

\def\be{\begin{equation}}
\def\ee{\end{equation}}

\begin{document}

{\title{CORRELATED ERRORS IN HIPPARCOS PARALLAXES TOWARDS THE PLEIADES AND THE HYADES}
\author{\bf 
Vijay K. Narayanan and
Andrew Gould \footnote{Alfred P.\ Sloan Foundation Fellow}
}
\affil{Department of Astronomy, The Ohio State University, Columbus, OH 43210;}
\affil{ Email: vijay,gould@astronomy.ohio-state.edu}

\bigskip
\bigskip
\bigskip
\centerline{\bf ABSTRACT}
We show that the errors in the \h parallaxes towards the Pleiades and the 
Hyades open clusters are spatially correlated over angular scales of 
$2$ to $3$ degrees, with an amplitude of up to $2$ mas.
This correlation is stronger than expected based on the analysis of the
\h catalog.
We predict the parallaxes of individual cluster members, $\pi_{\rm pm}$,
from their \h proper motions, assuming that all the cluster members 
move with the same space velocity.
We compare these parallaxes with their \h parallaxes, $\pi_{\rm Hip}$, 
and find that there are significant spatial correlations in the latter 
quantity.
We derive a distance modulus to the Pleiades of $5.58 \pm 0.18$ mag
from the gradient in the radial velocities of the Pleiades members
in the direction parallel to the proper motion of the cluster.
This value, derived using a geometric method, agrees very well with the
distance modulus of $5.60 \pm 0.04$ mag determined using the main-sequence
fitting technique, compared with the value of $5.33 \pm 0.06$ mag
inferred from the average of the \h parallaxes of the Pleiades members.
We show that the difference between the main-sequence fitting distance
and the Hipparcos parallax distance can arise from spatially correlated
errors in the Hipparcos parallaxes of  individual Pleiades members.
Although the \h parallax errors towards the Hyades are
spatially correlated in a manner similar to those of the Pleiades,
the center of the Hyades is located on a node of this spatial
structure.
Therefore, the parallax errors cancel out when the average distance 
is estimated, leading to a mean Hyades distance modulus that agrees with the 
pre-Hipparcos value.
We speculate that these spatial correlations are also responsible for the 
discrepant distances that are inferred using the mean \h parallaxes to
some open clusters, although an agreement between the mean \h parallax 
distance and the main-sequence fitting distance to other clusters 
does not necessarily preclude spatially correlated \h parallax errors.
Finally, we note that our conclusions are based on a purely geometric
method and do not rely on any models of stellar isochrones.

\medskip


\section{INTRODUCTION}

Trigonometric parallax is a fundamental method for measuring distances
to astronomical objects and is the first rung of the cosmic distance 
ladder.
It is a purely geometric technique, without the need for any ill-understood
empirical correlations between two physical quantities, one of which
is dependent on the distance and the other independent of distance.
The Hipparcos Space Astrometry Mission (\cite{esa97}) has derived accurate 
absolute trigonometric parallaxes for about 120,000 stars distributed 
all over the sky, and has produced the largest homogeneous all-sky
astrometric catalog to date.
The global systematic errors in the \h parallaxes are estimated
to be $ \la 0.1 $ mas, while the random errors in parallaxes of 
individual stars are typically on the order of 1 mas 
(\cite{arenou95}; \cite{arenou97}; \cite{lindegren95}).
However, the mean \h parallax distances to some open clusters are \diff 
from their distances inferred using other techniques
(\cite{mermio97hip}; \cite{robichon97}; \cite{vanleeuwen97}), 
suggesting that the true systematic errors may be an order of magnitude 
larger, at least on small angular scales (Pinsonneault et al. 1998, 
hereafter  PSSKH98).
In this paper, we estimate the level of the systematic errors in the \h 
parallaxes towards the \plei and the Hyades clusters by comparing 
for each of the cluster members, their \h parallax distances 
with their relative distances inferred from their \h proper motions, 
assuming that all the cluster members move with the same bulk velocity.
We first determine the distance to the Pleiades cluster using a variant 
of the moving cluster method and then present the evidence
for spatial correlations in the \h parallaxes towards both the Pleiades
and the Hyades.

The distances to the Hyades and the Pleiades are fundamental quantities
in establishing the absolute level of the main-sequence in the HR diagram,
and hence in estimating the distances to open clusters using the 
main-sequence fitting technique.
Thereby, they provide the first calibration points in the extragalactic 
distance scale.
Hence, it is imperative that these distances are firmly established
using techniques that require minimal assumptions.
While the \h astrometric catalog provides  straightforward
distance estimates to these clusters from the mean of the
parallaxes of the cluster members, there are surprising differences
between the mean \h parallax distances and the distances estimated 
using other techniques, for some open clusters including the Pleiades
(\cite{mermio97hip}; \cite{robichon97}).
In particular, the distance modulus to the Pleiades derived using the 
mean of the \h parallaxes is almost $0.3$ mag smaller 
than that derived using the main-sequence fitting technique 
(\cite{vanleeuwen97}), while there is no such discrepancy for the Hyades 
(\cite{perryman98}; PSSKH98).
A confirmation of this $15\%$ shorter distance to the Pleiades from the 
\h parallaxes has serious implications for our understanding 
of stellar evolution.
For example, if the Pleiades stars are in fact $0.3$ mag fainter 
than they were previously thought to be, there must be a population of 
sub-luminous zero-age main-sequence field stars in the solar neighborhood 
that has so far escaped detection (\cite{soderblom98}).

The difference in the distance estimates using the \h parallaxes
and using the main-sequence fitting method are much larger than what 
would be expected from incorrect metallicities, and this has 
led to an active search for alternate explanations.
These alternatives range from the ``Hyades anomaly'' (\cite{crawford75})
arising from a low Helium abundance of the Hyades (\cite{stromgren82})
which therefore affects the relative distance between the
Hyades and the Pleiades, to the ``fourth parameter'' effect which 
states that a fourth parameter is required, in addition to the age, 
the metallicity, and the Helium abundance, to adequately describe 
solar-type stars (\cite{alexander86}; \cite{nissen88}; 
see Mermilliod et al. 1997 for a review of explanations
invoking all these different effects).
PSSKH98 showed that an impossibly large Helium abundance $(Y = 0.37)$ is
required for the Pleiades stars to reconcile the shorter value of the 
Pleiades distance inferred from the \h parallaxes with the main-sequence 
fitting distance, and proposed a simpler explanation that there are
spatial correlations in the \h parallax errors on  small angular scales.
All these drastic consequences of a shorter distance to the Pleiades
mean that we need to independently check if the \h parallaxes
towards this cluster are free from any systematic errors, before
invoking alternate explanations for the ``failure'' of the 
main-sequence fitting technique.

Here, we compare  the \h parallax distances to the members of the Pleiades 
and the Hyades clusters with their distances computed using the moving 
cluster method.
This method assumes that all the cluster members move with the
same space velocity and that the velocity structure of the
cluster is not significantly affected by rotation.
Under this assumption, we can predict the distance (and hence the 
parallax) to each of the individual cluster members if we know the
common space velocity of the cluster.
We use a variant of the moving cluster method --- the radial-velocity 
gradient method, to compute the distance to the Pleiades using
simple geometrical considerations.
We use this distance to estimate the common space velocity of all
the Pleiades members and then predict the parallaxes
of individual Pleiades members.
We then compare these parallaxes with the \h parallaxes 
of the same stars.
This enables us to test the accuracy of the \h parallaxes on small scales,
in a manner that is independent of any stellar isochrones.
We extend this analysis to the Hyades cluster using the common cluster
space velocity determined by Narayanan \& Gould (1998, hereafter NG98)
The principal result of this paper is that the \h parallaxes towards both 
the clusters are correlated with position on scales of about $3^{\circ}$, 
with an amplitude of about $1$ to $2$ mas.
While it is well known that the errors in the \h parallaxes are correlated
over small angular scales (Lindegren 1988, 1989, \cite{lindegren97}; 
\cite{arenou97b}; \cite{vanleeuwen98}), we find that the correlation
is probably stronger than previous estimates.

The outline of this paper is as follows.
We explain the different variants of the moving cluster method in \S2.
We describe our selection of  Pleiades cluster members from the \h catalog 
and our estimate of the average proper motion of the cluster in \S3.
In \S4, we derive the distance to the Pleiades from the gradient in the 
radial velocities of its members, in the direction parallel to the 
the proper motion of the cluster.
We compare this distance with the mean \h parallax distance
and give our estimates of the systematic errors in \h parallaxes
towards the Pleiades in \S5.
In \S6, we show that the same type of systematic errors are also present
in the \h parallaxes towards the Hyades.
We present our conclusions in \S7.
This is the second paper in the series in which we compare the 
\h parallaxes of open clusters with independent distances derived using
geometrical techniques, the first being a check of the Hipparcos
systematics towards the Hyades (NG98).
We note that we will drop the usual conversion factor 
$A_{v} = 4.74047 {\rm \ km}\, {\rm yr}\, {\rm s}^{-1}$ 
from all our equations for the sake of clarity,
 leaving it to the reader to include it in the appropriate equations.
This is equivalent to adopting the units of 
${\rm AU}\, {\rm yr}^{-1}$ for the velocities, although we will still 
quote the numerical values of the velocities in 
$ {\rm \ km}\, {\rm s}^{-1}$.

\section{MOVING CLUSTER METHODS}

The fundamental requirement for using the moving cluster method to
estimate the distance to a stellar cluster is that all the stars in the 
cluster have the same space velocity (${\bf V}$) to within the velocity 
dispersion of the cluster.
The three observables of the cluster members, namely, their radial velocities 
($V_{r}$), their proper motion vectors ($\mbox{\boldmath $\mu$}$), and their
angular separations ($\mbox{\boldmath $\theta$}$) from a suitably defined 
cluster center, are to a good approximation related by,
\be
{\bf V}_{T} = {\bf V} - V_{r}{\bf \hat r},
\label{eqn:vtdef}
\ee
\be
\mbox{\boldmath $\mu$} = \frac{{\bf V}_{T}}{d},
\label{eqn:mudef}
\ee
\be
\delta{\bf V}_{T} = - V_{r}\mbox{\boldmath $\theta$},
\label{eqn:delvtdef}
\ee
\be
\delta\mu_{\bot}  = -\left( \frac{V_{r}}{d} \right) \theta_{\bot},
\label{eqn:delmuperpdef}
\ee
\be
\delta\mu_{\parallel}  = -\left( \frac{V_{r}}{d} \right)\theta_{\parallel} -
\left( \frac{\delta d}{d} \right){\mu_{\parallel}},
\label{eqn:delmuparldef}
\ee
\be
\delta V_{r} = (\mbox{\boldmath $\theta$} \cdot \mbox{\boldmath $\mu$})d = \theta_{\parallel} \mu_{\parallel}d =  \theta_{\parallel} V_{T},
\label{eqn:delvrdef}
\ee
where ${\bf V}_{T}$ is the transverse velocity of the cluster member
in the plane of the sky, $V_{T} = \vert {\bf V}_{T} \vert $, the subscripts
$\bot (\parallel)$ for the quantities $\mu$ and $\theta$ refer to the 
components of the respective vectors perpendicular (parallel) to the 
proper motion vector, and
 $\delta x$ is the difference in quantity $x$
($ x = {\bf V}_{T}, \mu_{\bot}, \mu_{\parallel}, d$)
between the individual member star and its average value at the centroid 
of the cluster sample.
Equations~(\ref{eqn:vtdef})-(\ref{eqn:delvrdef}) assume that
$\vert \mbox{\boldmath $\theta$} \vert \ll 1$ (the small angle approximation),
that $(\delta d/d) \ll 1$, that the velocity dispersion of the cluster is 
small compared to its mean space velocity, and that the velocity structure of 
the cluster is not significantly affected by rotation, expansion, shear, etc.
Equations~(\ref{eqn:delmuperpdef}),~(\ref{eqn:delmuparldef}), and 
~(\ref{eqn:delvrdef}) give three independent measures of the distance to the 
cluster center, and we can derive a more accurate distance to the cluster
by taking their weighted average.
This can be effectively accomplished using the statistical parallax formalism,
as explained by NG98.

The two variants of the moving cluster method that are currently
in use depending on the nature of the available data are:
\begin{description}
\item [{(1)}:] The convergent-point method: The proper motions of the 
individual cluster members are used to derive a convergent point on the sky.
This information is combined with the average radial velocity of the 
cluster center to derive its  distance using equation~(\ref{eqn:delmuperpdef}).
This method has been successfully applied to the Hyades cluster for a very
 long time (\cite{boss08}; \cite{schwan91}; \cite{perryman98}).
Moreover, if there is independent information from high precision photometry
about the relative distances between individual cluster members, 
equation~(\ref{eqn:delmuparldef}) can also be used to derive a more
precise estimate of the cluster distance (NG98).
\item [{(2)}:] The radial-velocity gradient method: The radial velocities of 
the individual cluster members can be used to measure the gradient 
in the radial velocity across the face of the cluster, in the direction 
parallel to the proper motion of the cluster.
This can be combined with an estimate of the average cluster proper motion,
to derive the cluster distance using equation~(\ref{eqn:delvrdef}).
This technique was first used by Thackeray (1967) to derive the convergent 
point of the Scorpio-Centaurus association.
It has since been applied to determine the distance to the Hyades cluster
(\cite{detweiler84}; \cite{gunn88}) and to determine the convergent
point of the Pleiades cluster by assuming a distance (\cite{rosvick92a}).
\end{description}

The three equations~(\ref{eqn:delmuperpdef}), (\ref{eqn:delmuparldef})
and (\ref{eqn:delvrdef}) yield independent measures of the distance
to the cluster with relative weights $W_{i} = N_{i}(d_{i}/\sigma_{i})^{2}$
where $d_{i}$ and $\sigma_{i}$, $(i=1,2,3)$ are the three distances and
distance errors, and $N_{i}$ is the  number of stars used to 
estimate the cluster distance by method $i$.
These weights are approximately given by
\be
W_{1}  = N\left< \frac{(\theta_{\bot}V_{r})^{2}}{(d\sigma_{\mu})^{2} +\sigma_{\rm clus}^{2}} \right>,
\label{eqn:w1def}
\ee
\be
W_{2}  = N\left< \frac{(\theta_{\parallel}V_{r})^{2}}{(d\sigma_{\mu})^{2} +\sigma_{\rm clus}^{2} +(\sigma_{d}\mu)^{2}} \right>,
\label{eqn:w2def}
\ee
\be
W_{3}  = N\left< \frac{(\theta_{\parallel}V_{T})^{2}}{\sigma_{r}^{2} +\sigma_{\rm clus}^{2}} \right>,
\label{eqn:w3def}
\ee
where $\sigma_{r}$ and $\sigma_{\mu}$ are the errors in the radial velocities
and the proper motion respectively, $\sigma_{d}$ is the uncertainty in the 
relative distance to individual cluster members,
 and $\sigma_{\rm clus}$ is the velocity dispersion of the cluster.
The weight $W_{1}$ corresponds to the classical convergent-point 
moving cluster method using individual proper motions 
[eq. (\ref{eqn:delmuperpdef})], while $W_{2}$ corresponds to the extension
of this method using photometry to estimate the relative distances between
the cluster members [eq. (\ref{eqn:delmuparldef})].
The weight $W_{3}$ corresponds to the radial-velocity gradient method 
described by equation~(\ref{eqn:delvrdef}).

For the purpose of illustration, we assume that for the Pleiades cluster,
$\sigma_{\rm clus} = 0.7$ ${\rm km}\, {\rm s}^{-1}$,
$d\sigma_{\mu} = 0.9$ ${\rm km}\, {\rm s}^{-1}$, 
$\sigma_{r} = 0.3$ ${\rm km}\, {\rm s}^{-1}$,
 $\sigma_{d} \mu = 0.9$ ${\rm km}\, {\rm s}^{-1}$, 
$\left< \theta_{\parallel}^{2} \right>  = 
\left< \theta_{\bot}^{2} \right>  \equiv
\left< \theta ^{2}\right>,$ 
$V_{r} = (1/5)V_{T} = 6$ ${\rm km}\, {\rm s}^{-1}$ and
$ N_{3} = 2N_{2} = 2N_{1} = 140$.
This leads to $W_{1} : W_{2} : W_{3} = 0.009 : 0.005 : 1.0$, which shows
that $99\%$ of the information about the Pleiades cluster distance is in
equation~(\ref{eqn:delvrdef}), i.e, in the radial-velocity gradient method.
We will therefore use only the radial-velocity gradient method in this paper.
This is in sharp contrast to the situation for the Hyades where the 
relative weights are in the ratio $1 : 0.33 : 0.50$, and hence most of the
distance information is in the classical convergent point method as extended
by NG98.

\section{MEMBERSHIP AND AVERAGE PROPER MOTION}

The procedure for determining the distance to the Pleiades from the radial
velocity gradient [eq. (\ref{eqn:delvrdef})] requires an accurate estimate of
the average proper motion of the cluster center in an inertial frame.
In this section, we explain our procedure for selecting Pleiades members from 
the Hipparcos catalog and our estimate of the location and the average
proper motion of the centroid of these members.

\subsection{\it Cluster Membership}
We begin by selecting all the stars from the \h catalog that are within 
$10^{\circ}$ of an approximate center of the Pleiades cluster and whose
proper motions are consistent with them being Pleiades members.
We assume an average radial velocity at the cluster center of
 $5$ kms$^{-1}$, an average proper motion of
$\mu_{\alpha} = 20$ mas$\,$yr$^{-1}$, $\mu_{\delta} = -45$ mas$\,$yr$^{-1}$, 
an average distance of $d = 132$ pc and an isotropic cluster velocity 
dispersion of $\sigma_{\rm clus} = 0.8$ kms$^{-1}$.
These values are only representative of the true values and are
as such only approximately correct, although we find that the
final list of cluster members is not  very sensitive to these values.
For each star $i$, we predict its proper motion 
$\mbox{\boldmath $\mu$}_{{\rm pred}, i}$
using equations~(\ref{eqn:vtdef}) and (\ref{eqn:mudef}) and compute the
quantity $\chi^{2}_{i}$ defined as
\be
\chi^{2}_{i} = \left< \Delta \mbox{\boldmath $\mu$}_{i} \vert {\bf C}_{i}^{-1} \vert \Delta \mbox{\boldmath $\mu$}_{i} \right>,
\label{eqn:chi2def}
\ee
where $\Delta \mbox{\boldmath $\mu$}_{i} = (\mbox{\boldmath $\mu$}_{{\rm Hip}, i} - \mbox{\boldmath $\mu$}_{{\rm pred}, i})$,  
$\mbox{\boldmath $\mu$}_{{\rm Hip}, i}$ is its \h proper motion,
and where  we have employed Dirac notation,
\be
\left< X \vert {\mathcal{O}} \vert Z \right> = \sum_{i,j}X_{i}{\mathcal{O}}_{ij}
Z_{j}.
\label{eqn:diracdef}
\ee
The covariance matrix $C_{i}$ is the sum of three terms:
(a) the covariance matrix of the Hipparcos proper motion,
(b) the isotropic velocity dispersion tensor of the cluster divided by the 
square of the mean distance of the cluster, $(\sigma_{\rm clus}/d)^{2}$, and
(c) a matrix of the form 
$\theta_{d}^{2}(\mbox{\boldmath $\mu$}^{T}\mbox{\boldmath $\mu$})_{{\rm pred},i}
$, where we adopt $\theta_{d} \equiv (\delta d/d) = 6\%$.
The third term accounts for a finite depth of the Pleiades cluster along
the radial direction and allows a Pleiades member to be located either
in front of or behind the assumed fiducial distance $d$.
We select all the stars with $\chi^{2}_{i} \leq 9$ (corresponding 
to $3\sigma$) to be candidate Pleiades  members.
This procedure selects a total of 81 Pleiades candidates from the \h 
catalog.
These include all but 12 of the 74 Pleiades candidate stars in the \h
Input Catalog.
The proper motions of these 12 stars (with Hipparcos IDs HIP 16119, 17026, 
17684, 17759, 17832, 18018, 18046, 18106, 18149, 18201, 18748, 19496)
differ widely from the average proper motion of the Pleiades, and they
are therefore most likely to be non-members.

We predict the parallax of each of these Pleiades candidates 
using their Hipparcos proper motions and the average space velocity of 
the cluster as,
\be
\pi_{{\rm pm},i} = \frac{\left< ({\bf V_{t}})_{i} \vert {\bf C}_{i}^{-1} \vert 
\mbox{\boldmath $\mu$}_{{\rm Hip},i} \right> }{\left< ({\bf V_{t}})_{i} \vert {\bf C}_{i}^{-1} \vert ({\bf V_{t}})_{i} \right>},
\label{eqn:pipm}
\ee
where 
$({\bf V_{t}})_{i} = {\bf V_{c}} - (\hat {\bf r}_{i} \cdot {\bf V_{c}})\hat{ \bf r}_{i} $ 
is the transverse velocity of the cluster in the plane of the sky at the
position of the star $i$, and the covariance matrix ${\bf C}_{i}$ is the sum
of the velocity dispersion tensor of the cluster divided by the 
square of the mean distance to the cluster and the covariance matrix of 
the Hipparcos proper motion of star $i$.
The error in $\pi_{{\rm pm},i}$ is equal to 
$\left< ({\bf V_{t}})_{i} \vert {\bf C}_{i}^{-1} \vert ({\bf V_{t}})_{i} \right
>^{1/2}.$
We use this parallax and the $V_{J}$ magnitude from Tycho photometry
to estimate the absolute magnitude (and the associated 
error) of each of these Pleiades candidates.

Figure 1 shows the color-magnitude diagram of all these Pleiades candidates.
There is an easily identifiable main sequence in the color range
$0 < (B-V)_{J} < 0.9$ and there are a few stars that clearly lie
either above or below this sequence even after accounting for their 
magnitude errors.
We adopt a color-magnitude relation
\be
M_{V} = 4 + 5.57\left[ \left( B-V \right)_{J} - 0.5 \right]
\ee
in the color range $0 < (B-V)_{J} < 0.9$ and accept all the stars
that lie within $0.4$ mag of this line as Pleiades members.
The observed color-magnitude relation is quite steep for $(B-V)_{J} < 0$
and does not show an unambiguous main sequence.
Therefore, we assume that all the stars with $(B-V)_{J} \leq 0$ are 
Pleiades members.
We also reject one star (HIP 16431) whose error in proper motion is
greater than $4$ mas$\,$yr$^{-1}$.
This algorithm selects a total of 65 stars as Pleiades members from the 
Hipparcos catalog.
These members are shown as solid circles in Figure 1, while the non-members
and plausible binary systems are represented by the open circles.
To summarize our selection of Pleiades members, we first select a 
total of $81$ candidates from the \h catalog whose proper motions are
consistent with them being Pleiades members.
We predict their parallaxes from their \h proper motions assuming that 
they have the same space velocity as the centroid of the Pleiades.
We then enforce a photometric cut where we accept as Pleiades members
only those $65$ candidates that lie close to the Pleiades main-sequence 
in the color-magnitude diagram.

\begin{figure}
\centerline{
\epsfxsize=\hsize
\epsfbox[18 144 592 718]{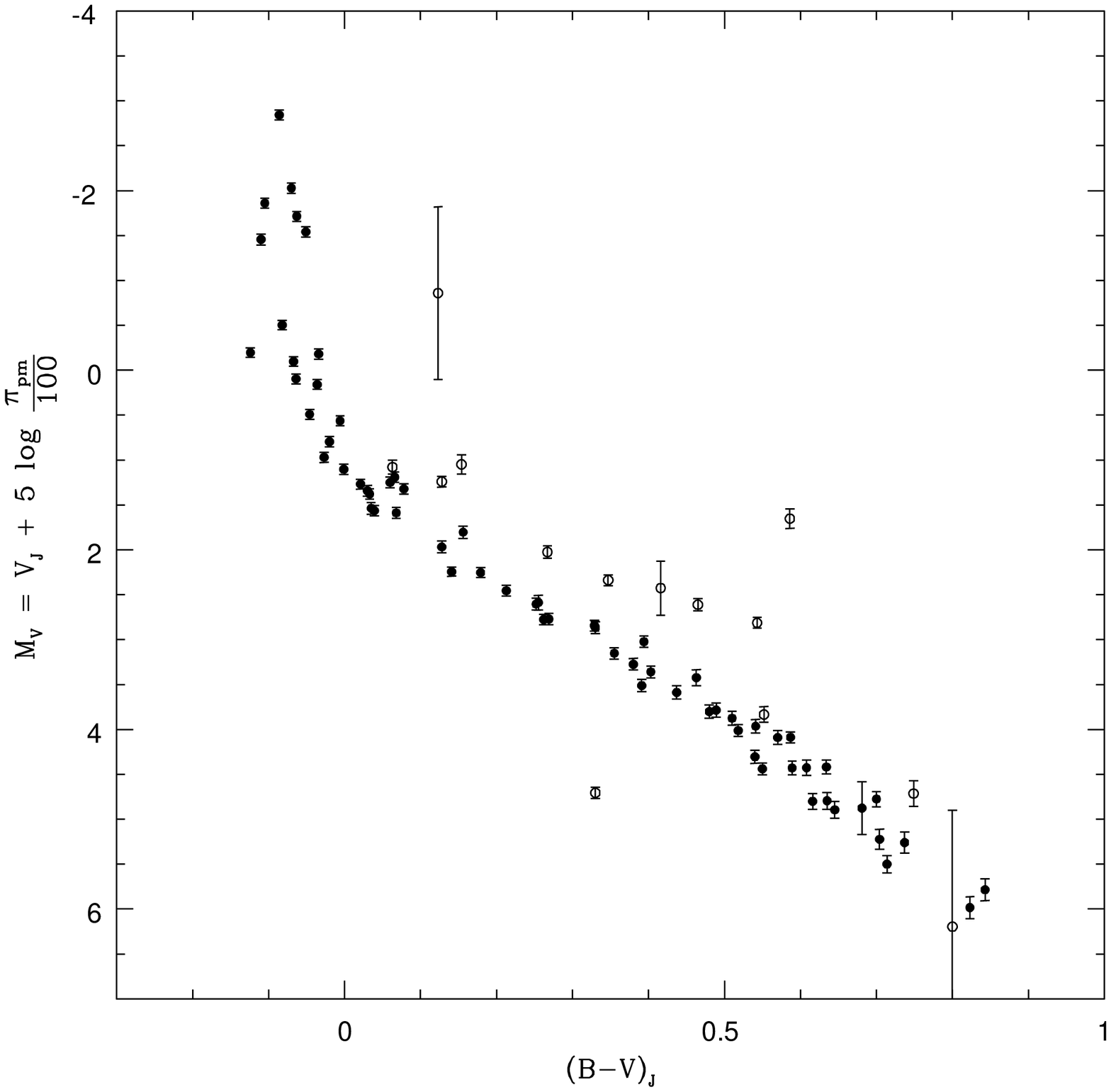}
}
\caption{ Color-magnitude diagram of all the stars in the \h catalog whose
individual proper motions are consistent with them being Pleiades members.
The parallax to each star is estimated from its \h proper motion, assuming a
common space velocity for all the Pleiades members.
The solid circles show the stars used to derive the average proper motion of
the Pleiades, while the open circles represent non-members and plausible 
binaries.
The colors and apparent magnitudes $(B-V)_{J}$ and $V_{J}$ are taken
from Tycho photometry.
}
\end{figure}

\subsection{\it Average proper motion}

We estimate the centroid and the average proper motion of the Pleiades 
cluster using all the 65 Pleiades members identified from the Hipparcos
catalog in \S3.1.
We compute the average proper motion at the cluster center as the mean
of all the individual proper motions of the Pleiades members weighted 
inversely by their covariance matrices.
The covariance matrix of each star is the sum of the covariance matrix
of the \h proper motions, the diagonal velocity dispersion tensor divided 
by the square of the mean distance of the cluster $(\sigma_{\rm clus}/d)^{2}$, 
and  a term arising from the distance ``dispersion'' 
$(\sigma_{d}/d)^{2})\mbox{\boldmath $\mu$}^{T} \cdot \mbox{\boldmath $\mu$}$,
to account for the non-zero depth of the Pleiades cluster.
The observed dispersion in the proper motions of the cluster members
in the direction perpendicular to the proper-motion vector includes
contributions from only the velocity dispersion term and the errors in the 
\h proper motions, while the observed dispersion parallel to the proper 
motion vector includes, in addition, a contribution from the dispersion
in the distances to individual Pleiades members.
Therefore, we estimate the dispersion in the proper motions from the 
difference between the observed and the \h proper-motion covariance matrices
in the perpendicular direction, and derive the distance ``dispersion''
as the difference between the observed covariance matrices in the parallel
and the perpendicular directions.

We find that the equatorial coordinate of the centroid of all the 65 
Pleiades members is $\alpha = 03^{h}46^{m}20^{s}, 
\delta = 23^{\circ}37.\hskip-2pt'0 $ (2000).
The average proper motion of the cluster 
at this location is 
${\overline \mu}_{\alpha} = 19.79 \pm 0.27$ mas$\,$yr$^{-1}$, 
${\overline \mu}_{\delta} = -45.39 \pm 0.29$ mas$\,$yr$^{-1}$, and the
correlation coefficient is $-0.087$.
Our estimate of the average proper motion of the Pleiades agrees well
with the estimate of
${\overline \mu}_{\alpha} = 19.67 \pm 0.24$ mas$\,$yr$^{-1}$, 
${\overline \mu}_{\delta} = -45.55 \pm 0.19$ mas$\,$yr$^{-1}$ 
by van Leeuwen \& Ruiz (1997).
We repeat the entire cluster-membership determination from the \h catalog stars
using this improved estimate of the average cluster proper motion and 
find that the membership does not change, showing 
that our selection of Pleiades members is not very sensitive 
to the initial values we have assumed for the average cluster proper motion.
Therefore, we will use these values for the average proper motion of the
Pleiades cluster in the remainder of this paper.

In our solution for the average proper motion of the Pleiades,
the dispersion in the proper motions is 
$(\sigma_{d}/d) = 1.63 \pm 0.38$ mas$\,$yr$^{-1}$.
Assuming a distance to the Pleiades of $d = 130.7$ pc (as we will find 
below), this dispersion in the proper motion corresponds to a velocity
dispersion of $1.00 \pm 0.24$ ${\rm km}\, {\rm s}^{-1}$, in reasonable
agreement with the value of $0.69 \pm 0.05$ ${\rm km}\, {\rm s}^{-1}$
we infer in \S4.2 from the radial velocities of the Pleiades members.
Similarly, we find a value of the distance ``dispersion'' of
$(\mu \sigma_{d}/d) = 1.37 \pm 0.74$ mas$\,$yr$^{-1}$ from the 
proper motions, corresponding to a depth of the cluster of
$(\sigma_{d}/d) = (2.77 \pm 1.49)\%$, which in angular scales is  
$\left< \theta_{d}^{2} \right>^{1/2} = 1^{\circ}\hskip-2pt.59 \pm 0^{\circ}\hskip-2pt.85$.
This is also in agreement with the angular dispersion 
of the $65$ cluster members in the directions perpendicular and parallel
to the average proper motion of the cluster, namely,
$\left< \theta_{\bot}^{2} \right>^{1/2} = 1^{\circ}\hskip-2pt.74 \pm 0^{\circ}\hskip-2pt.15$
and 
$\left< \theta_{\parallel}^{2} \right>^{1/2} = 2^{\circ}\hskip-2pt.03 \pm 0^{\circ}\hskip-2pt.18$.
Thus, the estimates of the cluster velocity dispersion
from both the proper motions and the radial velocities (which we will
estimate in \S4.2) are consistent with each other.
Similarly, the radial extent of the cluster that we infer from the proper 
motions is also comparable to the angular extent of the $65$ members of 
the Pleiades cluster.
We also find that $64$ of the $65$ Pleiades members are located 
within $6^{\circ}\hskip-2pt.2$ of the centroid of the cluster.

\section{RADIAL-VELOCITY GRADIENT AND CLUSTER DISTANCE}

We compute the distance to the Pleiades from the radial-velocity gradient 
method using the average proper motion derived in the previous section and the 
individual radial velocities of Pleiades members.
We now describe our selection of the Pleiades members with radial velocities
and our estimate of the distance to the cluster from its gradient in the
direction parallel to the average proper motion of the cluster.

\subsection{\it Radial velocity sample}

The Pleiades candidates in the \h catalog are mostly bright, early type 
stars with large rotational velocities.
Hence, it is difficult to measure their radial velocities from their spectra,
and the radial velocity surveys of Pleiades stars have been almost entirely
limited to faint, late type stars (later than the spectral type F).
Therefore, we select another list of fainter Pleiades members from the 
literature with measured radial velocities.

Our principal source of radial velocities is the radial-velocity survey of
the core and the corona stars in the Pleiades using the CORAVEL 
radial-velocity scanner
(Rosvick et al. 1992a, 1992b; \cite{mermio97}; \cite{raboud98}).
These three data sets contain the  radial-velocity data for
respectively, stars in the Pleiades corona selected on 
the basis of their proper motions and Walraven photometry by 
van Leeuwen, Alphenaar \& Brand (1986), stars in the outer
regions of the cluster selected on the basis of their proper motions 
by  Artyukhina \& Kalinina (1970), and stars in the inner region 
of the Pleiades in the Hertzsprung catalog (\cite{hertzsprung47}).
The radial velocities quoted in the three sources are the raw values 
measured from the spectra of these stars (J.C.\ Mermilliod 1998, 
private communication).
In practice, however, the measured radial velocities might include 
contributions from non-astrometric sources such as convective and 
gravitational line shifts, atmospheric pulsations etc. 
(\cite{dravins86}; \cite{nadeau88}).
The measured radial velocities must be corrected for all these
effects to estimate the true astrometric radial velocity of the stars.
However, these corrections are likely to be smaller than 
$1$ ${\rm km}\, {\rm s}^{-1}$ and therefore, we do not correct for 
these effects.
Further, it is possible that the three different sources of radial 
velocities have different zero-points, although this is unlikely to 
be a major problem for our sample of radial-velocity stars as all the 
radial velocities are measured using the same instrument.
We note here that our estimate of the distance to the Pleiades 
using the radial-velocity gradient method is insensitive to the absolute 
zero-point of the radial velocities, as long as it is the same for 
the three data sets.

We reject all the stars from these three datasets that are either known or
suspected to be binary systems, and which do not have any orbital solutions.
We include all the single stars and all the binary systems whose orbits
are either known from radial-velocity studies (\cite{mermio92}) or can be 
adequately constrained from infrared imaging (\cite{bouvier97}).
For the $9$ infrared binaries, we add an extra error in quadrature
of $\epsilon_{b} = \left[ M_{2}/\left(M_{1}+M_{2} \right) \right] \left[ G\left(M_{1} + M_{2} \right)/3a \right]^{1/2}$ to the quoted errors to reflect the 
uncertainty arising from the perturbative influence of the non-zero mass 
of the secondary stars (masses adopted from Bouvier, Rigaut \& Nadeau 1997),
and we accept only the $5$ stars with 
$\epsilon_{b} \leq 0.4 $ ${\rm km}\, {\rm s}^{-1}$ as Pleiades candidates.
Here, $M_{1}$ and $M_{2}$ are the masses of the primary and the secondary
stars, $a$ is the projected separation of the binary,
and the factor of $\sqrt{3}$ in the denominator is a fiducial
factor that roughly averages over all possible geometries of the 
binary orbits.
This procedure selects a total of $154$ Pleiades candidate stars with
measured radial velocities.

\subsection{\it Distance to the Pleiades}

Consider a cluster at a distance $d$, whose members all move with the 
{\it same} three space velocity, and let ${\bf n}$ be the direction vector 
towards the cluster center as defined by the sample used to compute the average
proper motion.
The observed radial velocity $V_{r, i}$ of any individual member star $i$
 located in the direction $ {\bf n}_{i}$ is related to the average 
radial velocity of the cluster center ${\overline V}_{r}$ by
\be
V_{r,i} = d \left( \mbox{\boldmath $\mu$} \cdot  {\bf n}_{i} \right) + {\overline V}_{r} \left(  {\bf n} \cdot  {\bf n}_{i} \right),
\label{eqn:vrdef}
\ee
where 
$\mbox{\boldmath $\mu$}$ is the average proper motion of the cluster.
This equation reduces to equation~(\ref{eqn:delvrdef}) under the small angle
approximation,
$\vert \mbox{\boldmath $\theta$} \vert \equiv \vert \cos^{-1} \left( {\bf n}_{i} \cdot {\bf n} \right) \vert \ll 1$, with
$\delta V_{r} \equiv \left( V_{r,i} - {\overline V}_{r} \right) $.
Since we determined $\mbox{\boldmath $\mu$}$ in \S3.2 for a sample of stars 
whose centroid is at $\alpha = 03^{h}46^{m}20^{s}, 
\delta = 23^{\circ}37.\hskip-2pt'0 $ (2000),
we must use the same direction for ${\bf n}$ in the present analysis, 
even though this is not the centroid of the radial-velocity sample.

We use equation~(\ref{eqn:vrdef}) to estimate the distance to the 
Pleiades ($d$) from the radial velocities of all the Pleiades candidates 
selected in \S4.1 
and the average cluster proper motion derived in \S3.2.
For each Pleiades candidate star $i$, we predict its radial velocity
$V_{r,i,{\rm pred}}$ at this cluster distance, and compute a quantity
$\chi_{v}^{2}$ defined
as
\be
\chi_{v}^{2} = \sum_{i=1}^{N} \frac {\left( V_{r,i} - V_{r,i,{\rm pred}} \right)^{2}}{\sigma_{v,i}^{2}},
\ee
where $\sigma_{v,i}$ is the sum in quadrature of the errors in the observed 
radial velocity of star $i$ and the velocity dispersion
of the cluster ($\sigma_{\rm clus}$), and $N$ is the number of Pleiades 
candidates.
We adjust the value of $\sigma_{\rm clus}$ so that the total value of 
$\chi_{v}^{2}$ is equal to $(N-2)$, 
and reject as non-members all the stars whose individual contributions
to $\chi_{v}^{2}$ is greater than 9 (corresponding to a $3\sigma$ outlier).
We repeat this procedure with the reduced list of candidates until there
are no stars whose individual contributions to $\chi_{v}^{2}$ are greater 
than 9.

We adopt as Pleiades members all the $141$ of the $154$ candidate stars that 
remain after the last iteration and derive a distance to the Pleiades of 
$d = 130.7 \pm 11.1$ pc, a velocity dispersion of $\sigma_{\rm clus} = 0.69 \pm 0.05$ ${\rm km}\, {\rm s}^{-1}$,
and a radial velocity of the centroid of the cluster of 
${\overline V}{r} = 5.74 \pm 0.07 $ ${\rm \ km}\, {\rm s}^{-1}$.
The total $\chi_{v}^{2}$ at the end of the last iteration is 
$139$ for a total of 141 stars, corresponding to $139$ degrees of
freedom.
The distribution of individual contributions to $\chi_{v}^{2}$ around the 
cut-off value of 9 are 
$6.1, 6.4, 8.0, 10.1, 12.2, 12.6, 25.4, 25.5, 49.9, 60.6 $, where
we include the first 3 stars with the values less than 9 as Pleiades members.
The individual contributions to $\chi_{v}^{2}$ are not distributed as the
square of a Gaussian, and there is a clear break in the distribution 
around $13$, although there is no clear break in the individual 
$\chi_{v, i}^{2}$ values at 9.
The three stars with individual $\chi_{v, i}^{2}$ in the range 
$9 < \chi_{v, i}^{2} < 12 $ are plausible members, while the stars
with $\chi_{v, i}^{2} > 20$ are most likely to be binary systems
or non-members.
We find that if we include these three plausible members,
the cluster distance is $d = 132.9 \pm 12.4$ pc, the new velocity dispersion 
is $\sigma_{\rm clus} = 0.80 \pm 0.07$ ${\rm km}\, {\rm s}^{-1}$, and the radial 
velocity of the centroid of the cluster is
${\overline V}_{r} = 5.80 \pm 0.08 $ ${\rm \ km}\, {\rm s}^{-1}$.
The total $\chi_{v}^{2}$ is $142$ for a total of 144 stars, corresponding 
to $142$ degrees of freedom.
This shows that our estimate of the cluster distance is not very sensitive
to the uncertainty in the cluster membership, and yields values around
$d = 130 $ pc as long as we reject the extreme outliers.

Figure 2 shows the radial-velocity difference
$[V_{r,i} - {\overline V}_{r} \left(  {\bf n} \cdot  {\bf n}_{i} \right)]
\simeq [V_{r,i} - {\overline V}_{r}]
$
for all the Pleiades candidates as a function
of the quantity ($ \mbox{\boldmath $\mu$} \cdot {\bf n}_{i}$).
The solid circles show the Pleiades members that are used to fit for the 
cluster distance, while the open circles represent the stars that are 
rejected as non-members by our algorithm.
The solid line shows our best-fit to the equation~(\ref{eqn:vrdef}),
and its slope is our estimate for the distance to the Pleiades.
We repeat here that the radial-velocity gradient method is 
a geometrical method, which relies on the assumption that the velocity 
structure of the Pleiades is not significantly affected by rotation.

\begin{figure}
\centerline{
\epsfxsize=\hsize
\epsfbox[18 144 592 718]{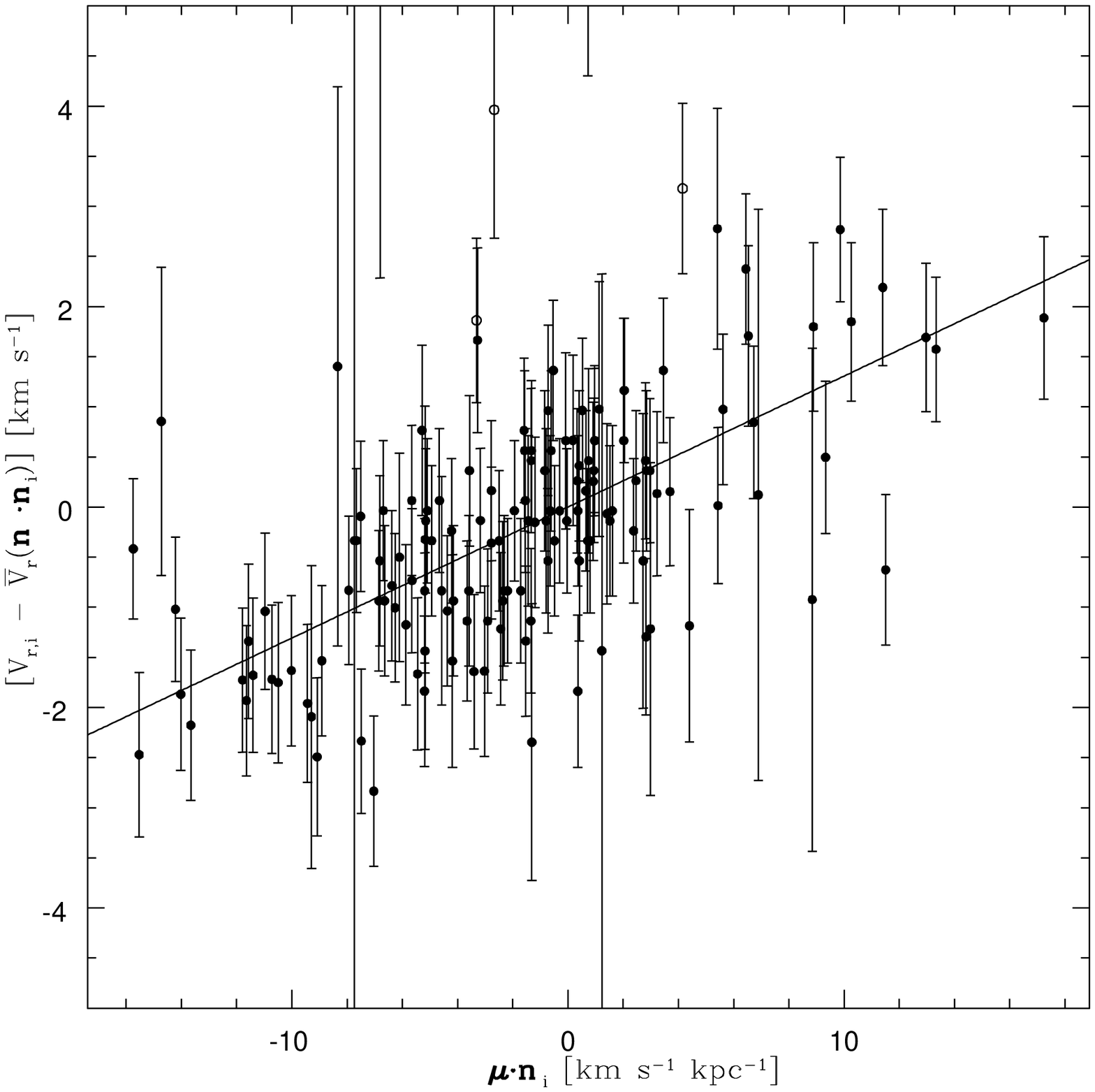}
}
\caption{Radial velocities of the Pleiades candidates as a function of the
scalar product of their mean Hipparcos proper-motion vector
($\mbox{\boldmath $\mu$}$) and 
the unit vector towards their position (${\bf n}_{i}$).
The slope of the best fit straight line is the distance to the Pleiades 
cluster, $d = 130.7 \pm 11.1$ pc.
The solid circles show the cluster members used to fit the straight line, 
while the open circles represent the stars that are rejected as non-members
by our algorithm.
}
\end{figure}

\section{COMPARISON WITH HIPPARCOS PARALLAXES}

The distance to the Pleiades from the radial-velocity gradient method
corresponds to a distance modulus of $(m-M) = 5.58 \pm 0.18$ mag.
This value agrees very well with the ``classical'' estimates of the
Pleiades distance modulus using main-sequence fitting techniques 
(\cite{vandenberg84}; \cite{eggen86}; \cite{vandenberg89}; \cite{pinsono98}),
all of which cluster around $5.60$ mag.
The discrepancy between the main-sequence fitting distance and the 
mean \h parallax distance to the Pleiades could arise for one of two reasons.
\begin{description}
\item [{(1)}] The \h parallaxes of the Pleiades members are 
systematically in error, and are larger on average than their true parallaxes.
\item [{(2)}] The isochrones that are used to derive the cluster 
distance  in the main-sequence fitting technique are all systematically 
too bright, leading to a larger distance for the Pleiades.
\end{description}
The theoretical isochrones are calibrated on the Sun
using accurate helioseismological data, and they are mostly used in a 
differential manner to derive the relative distances to clusters.
Furthermore, the distances to other open  clusters 
(e.g., the Hyades and $\alpha$ Per) using the same set of theoretical 
models are consistent with the \h parallax distances (\cite{pinsono98}).
Finally, only explanation (1) can account for the marginal discrepancy 
between the mean \h parallax distance to the Pleiades and 
the distance derived using the radial-velocity gradient method in \S4.
The distance modulus to the Pleiades using the rotational
modulation stars is also $5.60 \pm 0.16$ mag (\cite{odell94}), 
marginally larger than the mean \h parallax value and in very good 
agreement with the values from both the main-sequence fitting and 
the radial-velocity gradient techniques.
This consistency between the different independent methods of estimating the
distance to the Pleiades, all of which converge on a value of 
about $5.60$ mag, strongly suggests that there may be systematic errors
in the \h parallaxes towards the Pleiades.
We now extend our analysis to examine the spatial structure of these
errors.

Figure 3 shows the difference between $\pi_{\rm Hip}$, the Hipparcos 
parallaxes, and $\pi_{\rm pm}$, the parallaxes predicted using Hipparcos 
proper motions assuming that the members have a common space velocity, 
as a function of their angular distance from the centroid of the 
cluster $(\vert \theta \vert)$, for the $65$ Pleiades members that are
selected from the \h catalog using the procedure described in \S3.1.
The error bars show the quadrature sum of the errors in $\pi_{\rm Hip}$ and the
errors in $\pi_{\rm pm}$.
It is immediately obvious from this Figure that the Hipparcos parallaxes
are systematically larger than the parallaxes  predicted assuming common 
cluster motion, by up to $2$ mas, for all the stars that are located 
within $1^{\circ}$ of the centroid of the cluster.
The scatter in the values of $(\pi_{\rm Hip} - \pi_{\rm pm})$ increases for
$\vert \theta \vert > 1^{\circ}$, although it is clear that there is still
a systematic deviation from zero up to about $\vert \theta \vert = 2^{\circ}$.

\begin{figure}
\centerline{
\epsfxsize=\hsize
\epsfbox[18 144 592 718]{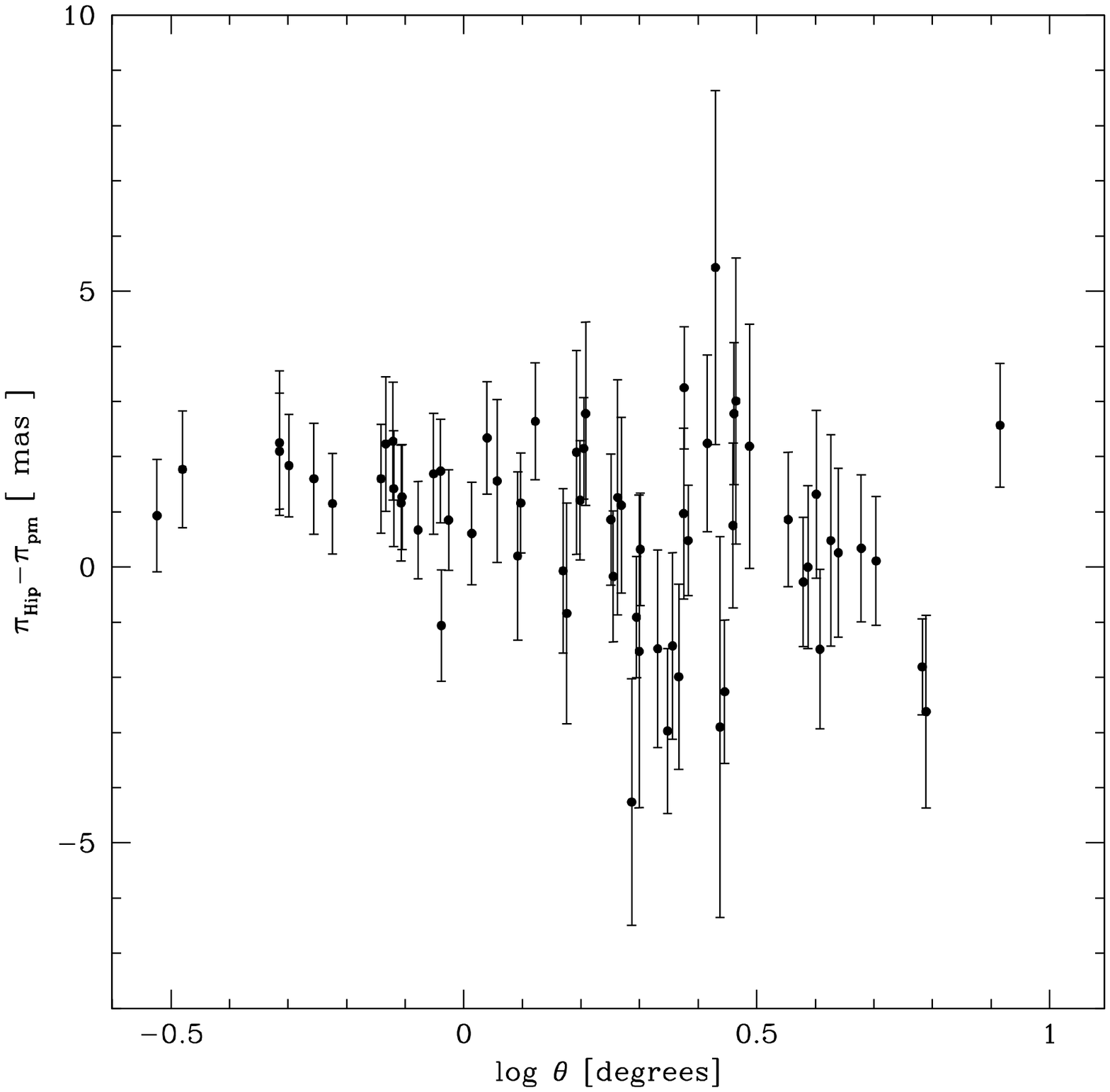}
}
\caption{ Difference between the Hipparcos parallaxes of individual stars,
$\pi_{\rm Hip}$, and their parallaxes predicted from their Hipparcos proper
motions assuming a common space velocity for all the cluster members,
$\pi_{\rm pm}$, as a function of the angular distance of the stars
from the centroid of the cluster 
($\theta = \vert \mbox{\boldmath $\theta$} \vert$).
The error bars show the quadrature sum of the errors in $\pi_{\rm Hip}$ and the
errors in $\pi_{\rm pm}$.
}
\end{figure}

Figure 4 shows the contours of the difference between the Hipparcos parallaxes 
$(\pi_{\rm Hip})_{s}$ smoothed on scales of $\theta_{s} = 1^{\circ}$
and the similarly smoothed parallaxes predicted from the Hipparcos proper
motions assuming a common space velocity for all the cluster members
$(\pi_{\rm pm})_{s}$, in an $8^{\circ} \times 8^{\circ}$ region about the 
centroid of the Pleiades cluster.
Solid contours correspond to $(\pi_{\rm Hip} - \pi_{\rm pm})_{s} \geq 0$, while
dashed contours correspond to $(\pi_{\rm Hip} - \pi_{\rm pm})_{s} < 0$.
The light contours range from $-1.8$ mas to $+2$ mas in steps of $0.1$ mas,
while the heavy contours range from $-1$ mas to $+2$ mas in steps of $1$ mas.
The solid circles show the positions of the individual Pleiades members.
We find this smoothed parallax difference field by computing the quantity
$(\pi_{\rm Hip} - \pi_{\rm pm})$ for each of the $65$ Pleiades members
and convolving this difference with a Gaussian filter 
$\exp(-\theta^{2}/2\theta^{2}_{s})/\sigma_{\rm tot}^{2}$, where
$\sigma_{\rm tot}^{2} = \sigma_{\rm Hip}^{2} + \sigma_{\rm pm}^{2}$.
The weighting by the inverse of the square of the error
ensures that the stars with noisy estimates of the parallax difference
are naturally given low weights when computing the smoothed parallax
difference field.
This Figure clearly shows that the \h parallaxes $\pi_{\rm Hip}$ are 
systematically larger than $\pi_{\rm pm}$ by up to $2$ mas, throughout the 
inner $6^{\circ} \times 6^{\circ}$ region around the centroid of the 
Pleiades.
Since very few of our $65$ cluster members are located outside 
the inner $4^{\circ} \times 6^{\circ}$ region, the smoothed field values
(the signal) outside this region comes primarily from the stars in the inner
region and therefore contains very little independent information
about the spatial structure of the systematic errors.
Hence, we restrict our quantitative analysis of this parallax difference
field of the Pleiades to the inner $4^{\circ} \times 6^{\circ}$ region
(shown by the dashed box in Fig. 4) in the remainder of this paper.

\begin{figure}
\centerline{
\epsfxsize=\hsize
\epsfbox[18 144 592 718]{f04.ps}
}
\caption{ Contours of the difference between the Hipparcos parallaxes 
$(\pi_{\rm Hip})_{s}$ smoothed on a scale of $\theta_{s} = 1^{\circ}$ 
and the similarly smoothed parallaxes predicted from the Hipparcos proper
motions assuming a common space velocity for all the Pleiades members
$(\pi_{\rm pm})_{s}$, in an $8^{\circ} \times 8^{\circ}$ region about 
the centroid of the Pleiades cluster.
Solid contours correspond to $(\pi_{\rm Hip} - \pi_{\rm pm})_{s} \geq 0$, while
dashed contours correspond to $(\pi_{\rm Hip} - \pi_{\rm pm})_{s} < 0$.
The light contours range from $-1.8$ mas to $+2$ mas in steps of $0.1$ mas,
while the heavy contours range from $-1$ mas to $+2$ mas in steps of $1$ mas.
The solid circles show the positions of the individual Pleiades members.
The dashed box shows the inner  $4^{\circ} \times 6^{\circ}$ region about 
the centroid of the Pleiades cluster.
}
\end{figure}

The spatial structure seen in Figure 4 can arise from spatially
correlated systematic errors in: (a) the Hipparcos parallaxes 
$\pi_{\rm Hip}$, or, (b) the parallaxes predicted from the \h proper motions
assuming a common space velocity for all the cluster members $\pi_{\rm pm}$,
or, (c)  both of these parallaxes.
Of these three possibilities, (a) will be true if there are as yet uncorrected
spatial correlations in the \h parallax errors on angular scales of a few 
degrees, while (b) will be the main source of error if the velocity field
of the Pleiades is dominated by substantial substructures that invalidate the 
assumption of a common space velocity for all the cluster members.
In principle, it is also possible that the structure arises from 
spatially correlated errors in the \h proper motions.
Indeed, if there are spatially correlated errors in \h 
parallaxes, it is reasonable to expect similar effects in the 
\h proper motions.
However, the structures seen in Figure 4 are of the same size 
($\sim 1$ mas)
as $\sigma_{\pi}({\rm Hip})$, the statistical errors in $\pi_{\rm Hip}$.
The statistical errors in $\pi_{\rm pm}$ arising from 
$\sigma_{\mu}({\rm Hip})$, the errors in the \h proper-motions, are 
smaller than this by a factor
$(\sigma_{\mu}/\mu)/(\sigma_{\pi}/\pi) \approx 1/6$.
Hence, one does not a priori expect correlations among the 
\h proper-motion  errors to have a noticeable effect.
Nevertheless, the tests that we carry out below would automatically
detect this unexpected effect if it were present.

To check which of the three alternatives is correct, we plot the 
quantities $(\pi_{\rm Hip} - \left< \pi_{\rm Hip} \right>)_{s}$  and 
$(\pi_{\rm pm} - \left< \pi_{\rm pm} \right>)_{s}$  in Figures 5 and 6 
respectively, in the same format as in Figure 4.
Here, $\left< \pi_{\rm Hip} \right> = 8.52 \pm 0.15$ mas and 
$\left< \pi_{\rm pm} \right> = 7.63 \pm 0.03$ mas are the average values,
computed using the $65$ Pleiades members,
of the \h parallaxes and  the parallaxes predicted assuming a 
common space velocity for all the cluster members.
The structures in Figure 5 closely resemble those in Figure 4 except
for a shift of the zero-point caused by the adoption of 
$\left< \pi_{\rm Hip} \right>$ as the Pleiades cluster parallax.
In Figure 6, on the other hand,  the inner $4^{\circ} \times 4^{\circ}$ region 
around the cluster center is remarkably smooth and close to zero and 
that there are no contours (either positive or negative) other than the 
one corresponding to $(\pi_{\rm pm} - \left< \pi_{\rm pm} \right>)_{s} = 0$.
This shows that the structures in $\pi_{\rm pm}$ arising from the 
errors in the \h proper motions are quite small compared to the 
structures arising from the correlations in the \h parallaxes.

\begin{figure}
\centerline{
\epsfxsize=\hsize
\epsfbox[18 144 592 718]{f05.ps}
}
\caption{ Contours of the difference between the smoothed Hipparcos 
parallaxes $(\pi_{\rm Hip})_{s}$ and the mean \h parallax of the $65$ Pleiades 
cluster members, $\left< \pi_{\rm Hip} \right>$.
Solid contours correspond to 
$(\pi_{\rm Hip} - \left< \pi_{\rm Hip} \right>)_{s} \geq 0$, while
dashed contours correspond to 
$(\pi_{\rm Hip} - \left< \pi_{\rm Hip} \right>)_{s} < 0$.
The light contours range from $-2$ mas to $+2.2$ mas in steps of $0.1$ mas,
while the heavy contours range from $-2$ mas to $+2$ mas in steps of $1$ mas.
The solid circles show the positions of the individual Pleiades members.
The dashed box shows the inner  $4^{\circ} \times 6^{\circ}$ region about 
the centroid of the Pleiades cluster.
}
\end{figure}

\begin{figure}
\centerline{
\epsfxsize=\hsize
\epsfbox[18 144 592 718]{f06.ps}
}
\caption{ Contours of the difference between the smoothed 
parallaxes predicted using the \h proper motions assuming a common
cluster space velocity for the members $(\pi_{\rm pm})_{s}$ and the 
mean value of this quantity for the $65$ Pleiades members,
$\left< \pi_{\rm pm} \right>$.
Solid contours correspond to 
$(\pi_{\rm pm} - \left< \pi_{\rm pm} \right>)_{s} \geq 0$, while
dashed contours correspond to 
$(\pi_{\rm pm} - \left< \pi_{\rm pm} \right>)_{s} < 0$.
The light contours range from $-0.1$ mas to $+0.7$ mas in steps of $0.1$ mas,
while the heavy line represents the contour corresponding to
$(\pi_{\rm pm} - \left< \pi_{\rm pm} \right>)_{s} = 0$.
The solid circles show the positions of the individual Pleiades members.
The dashed box shows the inner  $4^{\circ} \times 6^{\circ}$ region about 
the centroid of the Pleiades cluster.
}
\end{figure}

It is clear from Figures 5 and 6 that the spatial structure in
Figure 4 arises primarily from the spatial structure in the \h parallaxes.
The parallaxes in the entire region South-East of the centroid of the 
cluster are systematically too large by up to $2$ mas, while
there are no regions inside the inner $4^{\circ} \times 6^{\circ}$ region
 where the parallax difference is less than $-0.5$ mas.
It is clear from Figure 4 that an average of the \h parallaxes of stars 
lying in this region will be systematically larger, leading to  
an underestimate of the distance to the Pleiades.
We note here that the spatial structure seen in the 
$(\pi_{\rm Hip} - \pi_{\rm pm})_{s}$ field in Figure 4 is independent 
of our distance scale to the Pleiades itself.
Thus, if our estimate of the Pleiades space velocity is wrong, so that
all of our estimates of $\pi_{\rm pm}$ are systematically in error,
the absolute levels of the contours will change, while the spatial structure 
itself will remain the same.
A one-dimensional analog of our Figure 5 is Figure 20 of
PSSKH98, which plots the \h parallaxes of individual Pleiades members as
a function of their angular distance from the cluster center.

We see from the spatial structure in the smoothed field 
$(\pi_{\rm Hip} - \pi_{\rm pm})_{s}$ in Figure 4 that the \h parallax errors 
are correlated with position on angular scales of about $3^{\circ}$, with an 
amplitude of up to $2$ mas.
This is much larger than the upper limit of $0.1$ mas to the error in 
the global zero-point of the \h parallaxes 
(\cite{arenou95}; \cite{arenou97}), which, however, is valid only on
large angular scales.
Our estimate of the systematic errors demonstrates that they could be 
an order of magnitude larger than this on small angular scales, as was
already suggested by PKSSH98.

Even before the launch of the Hipparcos satellite, it was anticipated that
the errors in the \h parallaxes would be correlated over angular
scales of a few degrees (\cite{hoyer81}; Lindegren 1988, 1989).
The analysis of the \h parallaxes showed that the parallax errors are 
indeed strongly correlated on small scales, although the correlation 
becomes negligible for angular separations greater than about
$4^{\circ}$ (\cite{lindegren97}; \cite{arenou97b}; \cite{vanleeuwen98}).
An empirical fit to this correlation is given by the
function (\cite{lindegren97}):
\be
R(\theta) = R(0)\exp(-0.14\theta -1.04\theta^{2} +0.41\theta^{3} -0.06\theta^{4}),
\label{eqn:rthetadef}
\ee
where the angular separation, $\theta$, is measured in degrees, and 
$R(0) = 0.59$.
We now estimate how likely it is to get a parallax difference map
$(\pi_{\rm Hip} - \pi_{\rm pm})_{s}$ with the severe fluctuations seen in
Figure 4 if the errors in $\pi_{\rm Hip}$ are correlated according
to equation~(\ref{eqn:rthetadef}).

Figure 7 shows the normalized distribution of the fluctuation amplitude, $A$,
in the quantity $(\pi_{\rm Hip} - \pi_{\rm pm})_{s}$, if the errors in 
\h parallaxes are correlated over small angular scales as described 
by equation~(\ref{eqn:rthetadef}).
We define A as,
\be
A =  \left[ \left< (\pi_{\rm Hip} - \pi_{\rm pm})_{s}^{2} \right> - \left< (\pi_{\rm Hip} - \pi_{\rm pm})_{s} \right>^{2} \right]^{1/2}.
\label{eqn:adef}
\ee
We compute this distribution of A from an ensemble of $5000$ Monte-Carlo
realizations of the parallax differences $(\pi_{\rm Hip} - \pi_{\rm pm})_{s}$.
At each Monte-Carlo experiment, we assign a value of 
$(\pi_{\rm Hip, i} - \pi_{\rm pm, i})$ to each of the $65$ members that
is drawn from a Gaussian distribution whose variance is
$\sigma^{2}_{\rm tot, i} =  \sigma_{\pi, i}^{2}({\rm Hip}) + \sigma_{\pi, i}^{2}({\rm pm})$ and whose correlation with the other stars is described by 
equation~(\ref{eqn:rthetadef}).
We then compute A using only the values of the smoothed parallax difference 
field within the inner $4^{\circ} \times 6^{\circ}$ region of the centroid of 
the cluster.
The arrow in Figure 7 shows the value of the observed fluctuation amplitude
in the same region, $A_{\rm obs} = 0.47$ mas, for the field shown in Figure 4.
The probability of obtaining a fluctuation amplitude greater than the observed 
value is $P(A > A_{\rm obs}) = 17.7\%$, if the errors in the \h parallaxes
are correlated according to equation~(\ref{eqn:rthetadef}).

There is a small, but finite probability that the fluctuation amplitude
of the smoothed parallax differences $(\pi_{\rm Hip} - \pi_{\rm pm})_{s}$
towards the Pleiades is as high as that seen in Figure 4.
However, the modest probability of $17.7\%$ suggests that there might be 
angular correlations in the \h parallax errors over and the above the
correlation described by equation~(\ref{eqn:rthetadef}).
We now check to see if there is additional evidence for these extra 
correlations in the \h parallax errors towards the Hyades open cluster.

\begin{figure}
\centerline{
\epsfxsize=\hsize
\epsfbox[18 144 592 718]{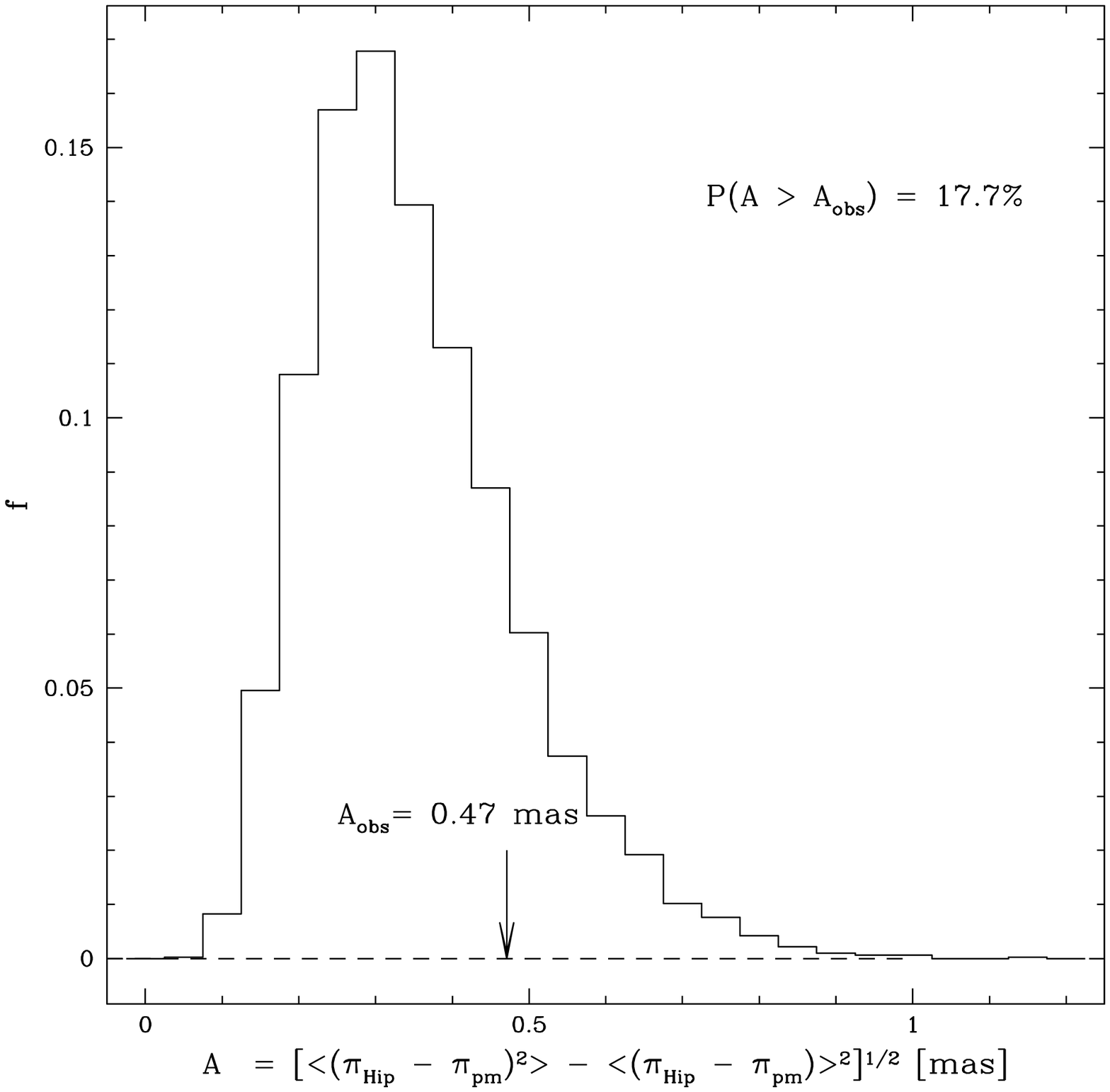}
}
\caption{Normalized distribution of the fluctuation amplitude, $A$, in the 
difference between the smoothed \h parallaxes $(\pi_{\rm Hip})_{s}$,
and the parallaxes predicted from the Hipparcos proper motions assuming
a common space velocity for all the Pleiades members $(\pi_{\rm pm})_{s}$, 
in a $4^{\circ} \times 6^{\circ}$ region 
about the center of the Pleiades cluster.
This distribution is computed assuming that the parallax
differences for each of the $i = 1, 2, \ldots  65$ Pleiades members
are distributed as a Gaussian whose variance is
$\sigma^{2}_{\rm tot, i} =  \sigma_{\pi, i}^{2}({\rm Hip}) + \sigma_{\pi, i}^{2}({\rm pm})$ and whose correlation with the other stars is 
described by equation~(\ref{eqn:rthetadef}).
The arrow shows the observed fluctuation amplitude in the same region,
$A_{\rm obs} = 0.47$ mas, for the field shown in Figure 4.
}
\end{figure}

\section{SYSTEMATIC ERRORS TOWARDS HYADES}

The analysis in the previous section shows that the \h parallax errors
towards the Pleiades cluster are spatially correlated over angular
scales of a few degrees, beyond what is expected from the 
analysis of the entire \h catalog.
We now check to see if these extra angular correlations are also present
in the \h parallax errors towards the Hyades.
If we do find extra correlations towards the Hyades, it is possible that 
these correlations are generic features of the \h parallax errors
all over the sky.
We describe our selection of Hyades members from the \h catalog
in \S6.1, and analyze the systematics of their  \h parallax errors
in \S6.2

\subsection{\it Hyades Membership}
We start by selecting a sample of stars from the \h catalog that are 
likely to be Hyades members based on their \h proper motions, using the 
procedure described in \S3.1.
We assume that the centroid of the Hyades cluster is at a distance
of $46.5$ pc towards the direction $\alpha = 04^{h}26^{m}32^{s},
\delta = 17^{\circ}13.\hskip-2pt'3 $ (2000), the velocity dispersion
of the cluster is $\sigma_{\rm clus} = 320 {\rm \ m}\, {\rm s}^{-1}$, and the bulk velocity
of the cluster in equatorial coordinates is
$(V_{x}, V_{y}, V_{z}) = (-5.41, 45.45, 5.74) {\rm \ km}\, {\rm s}^{-1}$,
as determined by NG98 using the statistical parallax algorithm.
For each star that is within $30^{\circ}$ of this direction,
we form the quantity $\chi_{i}^{2}$ as defined in 
equation~(\ref{eqn:chi2def})
 and select the stars whose $\chi_{i}^{2}$ is
less than $9$ to be Hyades candidates.
We adopt a value of the distance dispersion 
$\theta_{d} \equiv (\delta d/d) = 15\%$ to account 
for the finite depth of the cluster.
This procedure selects a total of $204$ Hyades candidates from the
\h catalog.
We use equation~(\ref{eqn:pipm}) to predict the parallaxes 
(and the associated errors) of these Hyades candidates from their \h 
proper motions, assuming that all the cluster members move with the same 
space velocity.
We estimate the absolute magnitudes of these stars using the
parallaxes derived in this manner and their apparent $V_{J}$ magnitudes from 
Tycho photometry.

Figure 8 shows the color-magnitude diagram of these Hyades candidates.
We have plotted only the $197$ candidates whose absolute magnitude errors 
are less than $1$ mag.
We see that there is an obvious main-sequence, and there are a few
stars lying above and below it.
These are most likely to be non-members.
We see that the main sequence in the color range $0.1 < (B-V)_{J} < 0.6$ 
has a steeper slope and a larger width compared to that in the 
color range $0.6 < (B-V)_{J} < 1.5$.
The larger width on the blue side probably arises from unidentified 
binary systems.
Accordingly, we fit different color-magnitude relations in each of 
these color ranges and select all the Hyades candidates that lie within
a finite width of these relations as Hyades members.
Our color magnitude relation for the Hyades is,
\be
M_{V} = \cases{ 2.72 + 7.14 \left[ \left( B-V \right)_{J} - 0.35 \right] 
	&if $0.1 < (B-V)_{J} < 0.6$ \cr
	6.44 + 4.84 \left[ \left( B-V \right)_{J} - 1.00 \right] 
	&if $0.6 < (B-V)_{J} < 1.5$. \cr}
\label{eqn:hyadescmd}
\ee
The solid line in the Figure shows this relation.
We assume that all the stars that lie within $0.4$ mag of the blue CMD
relation, or, within $0.24$ mag of the red CMD relation, are Hyades
members.
The $132$ Hyades members selected by this procedure are represented by 
the solid circles, while the non-members and plausible binaries are shown 
by the open circles.

\begin{figure}
\centerline{
\epsfxsize=\hsize
\epsfbox[18 144 592 718]{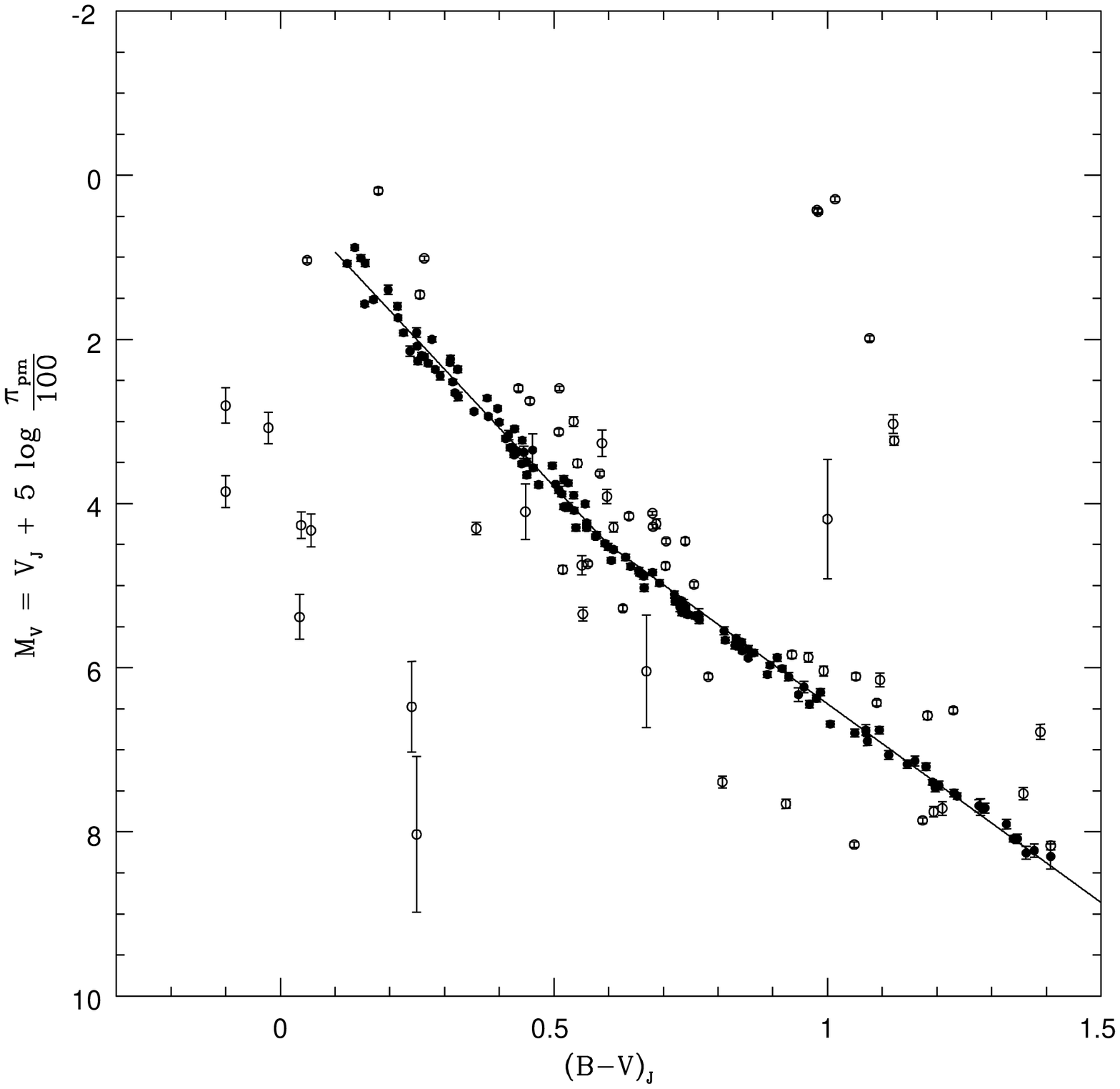}
}
\caption{ Color-magnitude diagram (CMD) of the $197$ stars in the \h catalog 
whose individual proper motions are consistent with them being Hyades members,
and whose absolute magnitude errors are smaller than $1$ mag.
The parallax to each star is estimated from its \h proper motion, assuming a
common space velocity for all the Hyades cluster members.
The solid circles show the stars that are most likely to be Hyades members
based on their location in the CMD, while the open circles represent 
non-members and plausible binaries.
The solid line shows our fit for the color-magnitude relation of the Hyades.
The colors and the apparent magnitudes $(B-V)_{J}$ and $V_{J}$ are taken
from Tycho photometry.
}
\end{figure}

\subsection{\it Systematics in Hipparcos parallaxes}
Figure 9 shows the contours of the difference between the Hipparcos parallaxes 
$(\pi_{\rm Hip})_{s}$ smoothed on scales of $\theta_{s} = 1^{\circ}$
and the similarly smoothed parallaxes predicted from the Hipparcos proper
motions assuming a common space velocity for all the cluster members
$(\pi_{\rm pm})_{s}$, in an $8^{\circ} \times 8^{\circ}$ region about the 
centroid of the Hyades cluster.
This Figure for the Hyades is analogous to Figure 4 for the Pleiades.
We find this smoothed parallax difference field using the $132$  Hyades
members, in the same manner as described in \S5 for the Pleiades.

The smoothed parallax difference field in Figure 9 clearly shows that the 
\h parallaxes $\pi_{\rm Hip}$ towards the Hyades are also spatially 
correlated over angular scales of a few degrees, with an amplitude of 
about $1$ to $2$ mas.
We have also plotted (but do not show) the quantities 
$(\pi_{\rm Hip} - \left< \pi_{\rm Hip} \right>)_{s}$  and 
$(\pi_{\rm pm} - \left< \pi_{\rm pm} \right>)_{s}$
for the Hyades, in a manner similar to Figures 5 and 6 for the Pleiades.
Once again, we find that the spatial structure in Figure 9 arises
from the structure in the \h parallaxes towards the Hyades and 
is not due to the structure in $(\pi_{\rm pm})_{s}$.
However, unlike the \h parallaxes towards the Pleiades, which were all
too large in the entire inner $4^{\circ}  \times 6^{\circ}$ region, 
the \h parallaxes towards the Hyades are systematically larger in 
some regions [e.g, a region of $2^{\circ} \times 2^{\circ}$ centered
on $(\Delta \alpha, \Delta \delta) = (+3^{\circ}, -1^{\circ})$], and 
systematically smaller in other regions [e.g, a region of 
$2^{\circ} \times 2^{\circ}$ centered on 
$(\Delta \alpha, \Delta \delta) = (-1^{\circ}, 1.5^{\circ})$].
Hence, the average value of the parallax difference is close
to zero, when it is computed using all the Hyades members that lie
in different regions.
This, combined with the large angular size of the Hyades cluster,
 can explain why the main-sequence fitting distance to the Hyades
agrees with the average of the \h parallaxes of its members (PSSKH98), 
although there  are significant spatial correlations in the 
\h parallax errors of the individual Hyades members.

\begin{figure}
\centerline{
\epsfxsize=\hsize
\epsfbox[18 144 592 718]{f09.ps}
}
\caption{ Contours of the difference between the Hipparcos parallaxes 
$(\pi_{\rm Hip})_{s}$ smoothed on a scale of $\theta_{s} = 1^{\circ}$ 
and the similarly smoothed parallaxes predicted from the Hipparcos proper
motions assuming a common space velocity for all the Hyades members
$(\pi_{\rm pm})_{s}$, in an $8^{\circ} \times 8^{\circ}$ region about 
the centroid of the Hyades cluster.
Solid contours correspond to $(\pi_{\rm Hip} - \pi_{\rm pm})_{s} \geq 0$, while
dashed contours correspond to $(\pi_{\rm Hip} - \pi_{\rm pm})_{s} < 0$.
The light contours range from $-1.4$ mas to $+1.4$ mas in steps of $0.1$ mas,
while the heavy contours range from $-1$ mas to $+1$ mas in steps of $1$ mas.
The solid circles show the positions of the individual Hyades members.
The dashed box shows the inner  $6^{\circ} \times 8^{\circ}$ region about 
the centroid of the Hyades cluster.
}
\end{figure}

Figure 10 shows the normalized distribution of the fluctuation amplitude, $A$,
in the quantity $(\pi_{\rm Hip} - \pi_{\rm pm})_{s}$, if the errors in 
\h parallaxes are correlated according to equation~(\ref{eqn:rthetadef}).
We compute this distribution in the same manner as described for the
Pleiades cluster.
We compute the fluctuation amplitude only within the 
inner $6^{\circ} \times 8^{\circ}$ region (shown by the dashed box in 
Figure 9) around the centroid of the Hyades.
The arrow in Figure 10 shows the value of the observed fluctuation amplitude
in the same region, $A_{\rm obs} = 0.62$ mas, for the field shown in Figure 9.
The probability of obtaining a fluctuation amplitude greater than the observed 
value is $P(A > A_{\rm obs}) = 9.1\%$.

We see that there is a only modest probability of obtaining a fluctuation 
amplitude that is as large as the observed value.
This is similar to the case of the Pleiades, although the probability 
in the case of the Hyades is almost a factor of two smaller than that
for the Pleiades.
The joint probability of obtaining the observed fluctuation amplitudes
for both the Pleiades {\it and} the Hyades is only about $1.6\%$, if the 
smoothed parallax differences of the Pleiades and the Hyades clusters 
are independent random processes.
This supports our speculation that there might be stronger angular 
correlations in the \h parallax errors, beyond the model described
by equation~(\ref{eqn:rthetadef})

\begin{figure}
\centerline{
\epsfxsize=\hsize
\epsfbox[18 144 592 718]{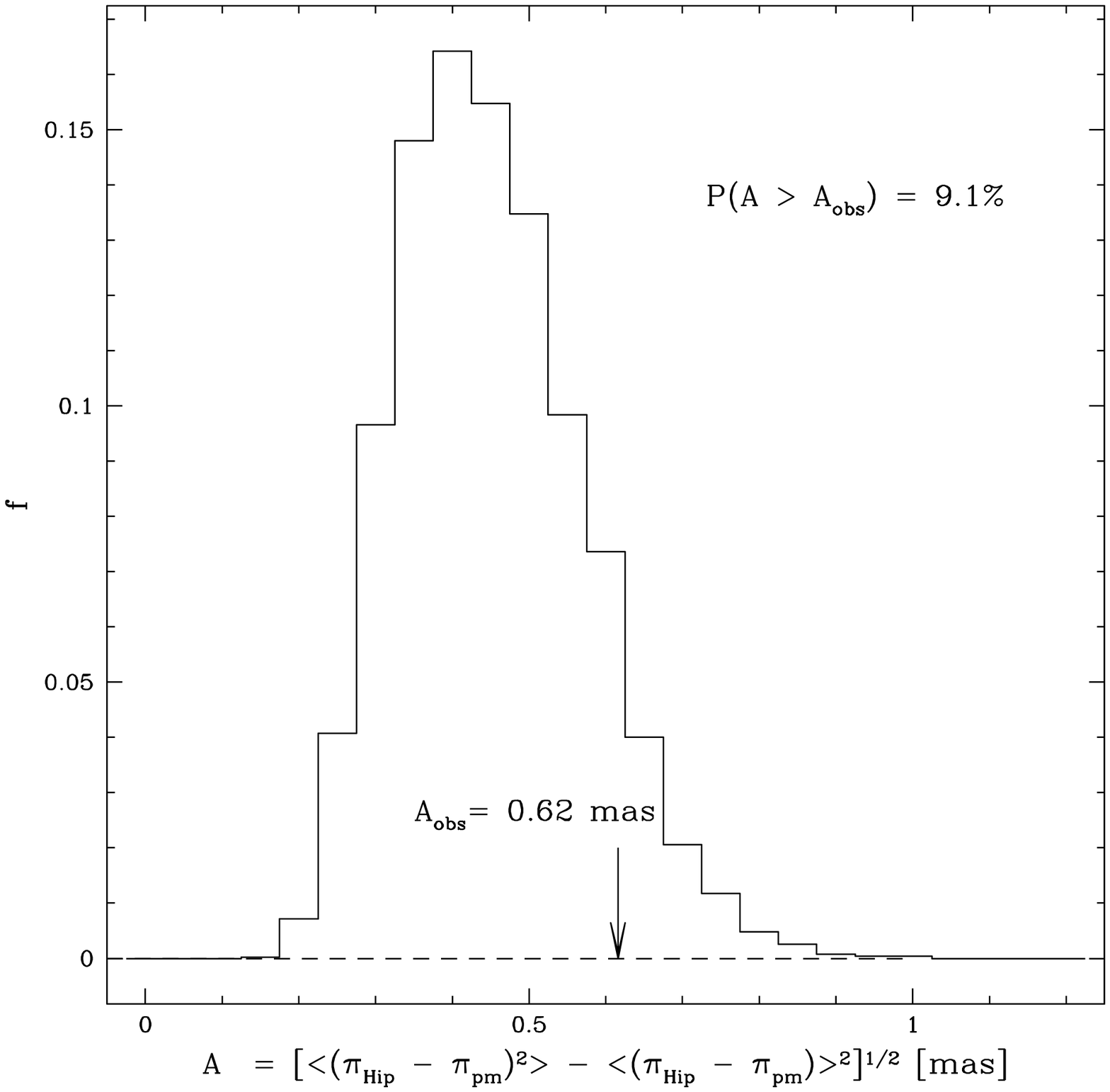}
}
\caption{Normalized distribution of the fluctuation amplitude, $A$, in the 
difference between the smoothed \h parallaxes $(\pi_{\rm Hip})_{s}$,
and the parallaxes predicted from Hipparcos proper motions assuming
a common space velocity for all the Hyades members $(\pi_{\rm pm})_{s}$, 
in a $6^{\circ} \times 8^{\circ}$ region 
about the center of the Hyades cluster.
This distribution is computed  assuming that the parallax
differences for each of the $i = 1, 2, \ldots  132$ Hyades members
are distributed as a Gaussian whose variance is
$\sigma^{2}_{\rm tot, i} =  \sigma_{\pi, i}^{2}({\rm Hip}) + \sigma_{\pi, i}^{2}({\rm pm})$ and whose correlation with the other stars is described by 
equation~(\ref{eqn:rthetadef}).
The arrow shows the observed fluctuation amplitude 
in the same region, $A_{\rm obs} = 0.62$ mas, for the field shown in Figure 9.
}
\end{figure}

\section{CONCLUSIONS}

The \h mission has derived absolute trigonometric parallaxes to about 
$120,000$ stars distributed all over the sky.
It is the largest homogeneous all-sky source of absolute parallaxes to date
and can potentially influence many branches of astronomy 
(see the review by Kovalevsky 1998).
Therefore, it is crucial to understand the errors in the \h astrometry.
Motivated by the increasing evidence that the distances to some open clusters
inferred from the mean \h parallaxes of their members are in conflict 
with their pre-Hipparcos values, we have critically analyzed the spatial 
correlations of the \h parallax errors on small scales.
Specifically, we have compared the \h parallaxes of the Pleiades and the 
Hyades cluster members with their parallaxes predicted from their
\h proper motions, assuming that all the cluster members  move with
the same space velocity.

Our main conclusions are as follows:
\begin{description}
\item [{(1)}:] We have derived a distance modulus to the Pleiades
of $(m-M) = 5.58 \pm 0.18$ mag using a variant of the moving
cluster method -- the gradient in the radial velocity of the cluster 
members in the direction of the proper motion of the cluster.
This value agrees very well with the distance modulus of 
$5.60 \pm 0.04$ mag derived using the classical main-sequence 
fitting technique (\cite{vandenberg89}; \cite{pinsono98}), but it is in
marginal conflict with the shorter distance modulus of $5.33 \pm 0.06$ mag
inferred by averaging the \h parallaxes of Pleiades members
(\cite{vanleeuwen97}).
The radial-velocity gradient method to estimate the cluster distance is a 
geometrical technique which relies on the assumption that the 
velocity structure of the Pleiades is not significantly affected by rotation.
\item [{(2)}:] We find that the \h parallax errors towards the Pleiades
cluster are spatially correlated over angular scales of 
2 to 3 degrees, with an amplitude of up to $2$ mas.
This can explain why the distance to the Pleiades cluster inferred
by  averaging the \h parallaxes of its members is smaller than 
its distance inferred by other techniques.
Even if the velocity distribution of the Pleiades members do
not conform to a common bulk space motion, we still see the spatial 
correlations in the \h parallaxes.
However, we cannot determine the zero-point of these fluctuations 
without the independent estimate of the cluster distance that comes
from the assumption of a common space velocity for all the 
cluster members (or some other parallax-independent source).
\item [{(3)}:] The spatial correlations in the \h parallaxes are also 
seen towards the Hyades cluster.
However, there are both positive and negative fluctuations in the \h 
parallax errors towards the region of the Hyades, with the result 
that these fluctuations cancel out on average and the distance to 
the Hyades inferred by averaging the \h parallaxes of all its members 
agrees well with other distance measurements.
\item [{(4)}:] The probabilities of obtaining the observed fluctuation 
amplitudes, $A_{\rm obs}$, in the smoothed parallax difference field 
$(\pi_{\rm Hip} - \pi_{\rm Hip})_{s}$, are small for both 
the Pleiades and the Hyades ($17.7\%$ and $9.1 \%$, respectively),
if the angular correlations in the \h parallax errors are described
by equation~(\ref{eqn:rthetadef}).
This suggests that there are almost certainly stronger spatial 
correlations in the \h parallax errors beyond what is modeled by 
equation~(\ref{eqn:rthetadef}).
Since we see these stronger correlations in \h parallax errors
towards both the Pleiades and the Hyades, we suggest that this may be  
a generic feature of the \h parallax errors all over the sky.
\end{description}

It is clear from the above conclusions that it is necessary to adopt a 
cautious approach when averaging the \h parallaxes over small angular scales.
In particular, it is necessary to quantify the effect of
spatial correlations in the parallaxes when dealing with a
distribution of stars that are separated by a few degrees.
Thus, for example, it has been found that when \h parallaxes are used to 
estimate the absolute magnitudes of stars in open clusters, such disparate 
open clusters as Praesepe, Coma Ber, $\alpha $ Per and Blanco I define the 
same main-sequence despite their widely different metallicities, 
with [Fe/H] ranging from $-0.07$ dex for Coma Ber to about $+0.23$ dex 
for Blanco I (\cite{mermio97hip}; \cite{robichon97}).
Our analysis shows that such an effect could arise from  
spatially correlated \h parallaxes of the cluster members, of the type
seen towards the Pleiades and the Hyades clusters.
Thus, a metal-rich cluster whose \h parallaxes are all systematically
larger than the true values 
can have the same apparent main-sequence as a metal-poor cluster whose 
systematic errors in different regions of the cluster 
cancel out on an average.
The discrepancy between the distances inferred from the 
average \h parallax and that inferred from the main-sequence fitting 
technique for other open clusters (e.g, for Coma Ber, PSSKH98)
could also arise from correlated parallax errors that do not cancel out on 
average, similar to the  situation in the Pleiades.
On the other hand, as we showed for the Hyades, an agreement
between these two distance measurements does not necessarily preclude 
stronger spatial correlations in the \h parallaxes.

Our work shows that there are strong spatial correlations in the 
errors of the parallaxes  in the \h catalog.
We note that this is not necessarily in conflict with the upper limit
of $0.1$ mas to the error in the global zero-point of the \h 
parallaxes over the full sky (\cite{arenou95}; \cite{arenou97}).
The global tests have very little power to probe for
systematic errors on smaller scales.
Finally, we note that, given the sparse average density of about 
$3$ stars$/\Box^{\circ}$ in the \h catalog, the open clusters with 
a large local concentration of stars may be the only regions where we can
test the small scale systematics in the \h catalog.

After the completion of this work, we became aware of the work of
van Leeuwen (1999), who has suggested the existence of an age-luminosity 
relation for main-sequence stars, in strong contradiction with the standard 
theory of stellar evolution.
Alternatively, if the small-angle correlations in the Hipparcos
parallaxes towards the Pleiades and the Hyades that we found in this
paper are a generic feature of Hipparcos parallaxes, then this proposed 
age-luminosity relation could be an artifact arising from an inadequate
treatment of these correlations.
A more detailed dicussion of this issue is beyond the scope of this paper
and will be addressed in the ongoing work of 
Pinsonneault et al. (1999, in preparation).

\acknowledgments

This work was supported in part by the grant AST 97-27520 from the NSF.
We thank Marc Pinsonneault, Bob Hanson, John Stauffer and Frederic Arenou
for helpful suggestions.
We also thank David Weinberg for his comments on an earlier draft 
of this paper.

{}
\end{document}